\documentclass[10pt]{bmc_article}

\usepackage{cite} 
\usepackage{url}  
\usepackage{ifthen}  
\usepackage{multicol}   
\usepackage[utf8]{inputenc} 
\usepackage{amsmath,amsfonts} 
\usepackage{graphicx}
\newboolean{publ}

\newcommand{\be}{\begin{eqnarray}}
\newcommand{\ee}{\end{eqnarray}}
\newcommand{\ba}{\begin{array}}
\newcommand{\ea}{\end{array}}
\newcommand{\bv}{\begin{verbatim}}
\newcommand{\ev}{\end{verbatim}}
\newcommand{\bt}{\begin{tabular}}
\newcommand{\et}{\end{tabular}}
\newcommand{\btab}{\begin{table}}
\newcommand{\etab}{\end{table}}
\newcommand{\bfig}{\begin{figure}}
\newcommand{\efig}{\end{figure}}
\newcommand{\bc}{\begin{center}}
\newcommand{\ec}{\end{center}}
\newcommand{\bit}{\begin{itemize}}
\newcommand{\eit}{\end{itemize}}

\newcommand{\Sum}{\displaystyle\sum\limits}
\newcommand{\tw}{\textwidth}
\newcommand{\ig}[1]{\includegraphics[width=#1\tw]}

\newcommand{\syn}{{\it Synechocystis sp PCC. 6803 }}

\newenvironment{bmcformat}{\baselineskip20pt\sloppy\setboolean{publ}{false}}{\baselineskip20pt\sloppy}

\begin{document}
\begin{bmcformat}

\title{PyNetMet: Python tools for efficient work with networks and metabolic models}

\author{D.~Gamermann\correspondingauthor$^{1,2}$%
         \email{Daniel Gamermann\correspondingauthor - daniel.gamermann@ucv.es}
       \and 
       A.~Montagud$^2$
       \and
       R.~A.~Jaime Infante$^3$
       \and
       J.~Triana$^3$
       \and
        P.~F.~de C\'ordoba$^2$
       and
        J.~F.~Urchuegu\'ia$^2$
      }

\address{\flushleft
    \iid(1)C\'atedra Energesis de Tecnolog\'ia Interdisciplinar, Universidad Cat\'olica de Valencia San Vicente M\'artir, \\ Guillem de Castro 94, E-46003, Valencia, Spain.\\
    \iid(2)Instituto Universitario de Matem\'atica Pura y Aplicada, Universidad Polit\'ecnica de Valencia, \\  Camino de Vera 14, 46022 Valencia, Spain. \\
    \iid(3)Universidad Pinar del R\'io ``Hermanos Sa\'iz Montes de Oca'',\\ Mart\'i 270, 20100, Pinar del R\'io, Cuba.
}

\maketitle

\begin{abstract}
{\bf Background:} The study of genome-scale metabolic models and their underlying networks is one of the most important fields in systems biology. The complexity of these models and their description makes the use of computational tools an essential element in their research. Therefore there is a strong need of efficient and versatile computational tools for the research in this area.

{\bf Results:} In this manuscript we present PyNetMet, a Python library of tools to work with networks and metabolic models. These are open-source free tools for use in a Python platform, which adds considerably versatility to them when compared with their desktop software similars. On the other hand these tools allow one to work with different standards of metabolic models (OptGene and SBML) and the fact that they are programmed in Python opens the possibility of efficient integration with any other already existing Python tool.

{\bf Conclusions:} PyNetMet is, therefore, a collection of computational tools that will facilitate the research work with metabolic models and networks.
\end{abstract}

\ifthenelse{\boolean{publ}}{\begin{multicols}{2}}{}

\section{Background}

Nowadays, the genome-scale reconstruction of metabolic models has become one of the corner stones of systems biology. Reconstructed metabolic models have been used in a wide range of applications, such as the study of metabolism regulation and operations \cite{r1, r2, r3, r4}, determination of the optimal conditions for the growth and prediction of maximum yield of biomass for a determined organism \cite{j1}, the search of potential sites for metabolic engineering \cite{en1}, the production of biofuels \cite{bf1, bf2, bf3} and even in the reconstruction of phylogenetic trees \cite{fgt1}. One of the most important computational tools for the analysis of metabolic models is the flux balance analysis (FBA) \cite{fba1}, which consists basically in the determination of a possible consistent solution for all fluxes in the reactions of the model that optimizes some given objective.

A particular way to study genome-scale metabolic models is to analyze their underling networks. The simplest example of such network is to define each metabolite present in a metabolism as a network node, and assign connections in between the nodes based on the connection of the respective metabolites in the metabolism through chemical reactions. Such networks have been widely studied in the literature \cite{nw1, nw2, nw3}.

Typical genome-scale metabolic models comprise around thousand different metabolites and chemical reactions and, correspondingly, the underlying metabolic networks are complex structures with around one thousand interconnected nodes. The analysis of these complex structures would be nearly unfeasible without the aid of modern computers. There are different available software for performing FBA on a metabolic model like the COBRA toolbox, originally developed for MatLab \cite{cobra}, but now also available for Python, or the OptFlux software \cite{optgene}, among others and also software for the analysis of networks. All these software have drawbacks. For instance, there are two different standards for the storage of metabolic models: the SBML \cite{hucka} and OptGene (also known as BioOpt) \cite{j2} formats, and the available software either use one or another, but not both. On the other hand, some software are not free (like MatLab) or are desktop software which limits their uses.

In this article we present a series of tools, which have been developed in Python, for dealing with chemical reactions and analyzing networks and metabolic models. Python is a free, open-source, modular, object oriented programming language \cite{python}. Open-source libraries boost the development and advance of bioinformatics by allowing developers and researchers to develop new tools and applications over modules already built. Moreover, modular programming languages like Python allow easy and efficient integration of its modules with other libraries and software (which is hardly done with desktop applications). Python has also available the Biopython package which already contains various standards used in bioinformatics and allows the direct connection with different biological databases.

The package present here is called PyNetMet (from Python Network Metabolism), it comprises four classes called \texttt{Enzyme}, \texttt{Network}, \texttt{Metabolism} and \texttt{FBA}. The \texttt{Enzyme} class defines a new object called \texttt{Enzyme} which stores a chemical reaction. The methods in this class will be thoroughly used in by the class \texttt{Metabolism} in order to organize and extract information from the list of chemical reactions that define any particular metabolic model. The class \texttt{Network} provides tools to study any graph defined by interconnected nodes. Several classical graph theory algorithms are programmed inside this class, like the Dijkstra's algorithm \cite{dij} for calculating the shortest mean distance between the nodes of the network or paths connecting any two nodes, the Kruskal algorithm \cite{krusk} to organize the nodes according to their clustering similitude, and others. The \texttt{Metabolism} class has basically two functions. First, it works as a parser from OptGene or SBML file formats, which can store metabolic models and the parameters needed for flux balance analysis (FBA). Secondly, it extracts and resumes information from the metabolism, allowing one to find disconnected components in the model, reactions that cannot contribute to the simulation of an organism's metabolism (FBA), among other tools. Finally the class \texttt{FBA} allows one to perform an FBA of the model, and apply other algorithms to search for essential reactions or calculate the sensitivity of the objective flux with respect to the flux in any reaction. PyNetMet can be downloaded from the Python Package Index (pypi.python.org/pypi/PyNetMet/1.0).


\section{Implementation}

The package PyNetMet consists of four classes: \texttt{Enzyme}, \texttt{Network}, \texttt{Metabolism} and \texttt{FBA}, all fully programed in Python 2.7 language. The class \texttt{Enzyme} has no dependencies, it defines a new type of variable which stores a single chemical reaction. Class \texttt{Network} has a single dependency (for two specific functions) which is the Python Imaging Package (PIL), for making plots representing the clustering of nodes. The Class \texttt{Metabolism} depends on the classes \texttt{Enzyme} and \texttt{Network}, and class \texttt{FBA} depends on the class \texttt{Metabolism} and on the Python library Pyglpk (which contains tools for solving the associated optimization linear problem).

In Tables \ref{tab1} and \ref{tab2} we list all attributes and methods (not the underscore ones), respectively, of the classes with a short description. For a more complete description and more detailed examples of use, please refer to the manual that accompanies the PyNetMet distribution and is available here as additional file 1. In what follows in this section, we present each class separately commenting on some important aspects and definitions. 

To use each class, one only has to import it as a Python module. A few examples will be given.

\subsection{Enzyme}

The class \texttt{Enzyme} defines an chemical reaction object. It will be the main object used to build the \texttt{Metabolism} object later on. Its obligatory input is a string containing the reaction. This string must be a reaction written in OptGene format, so it should have two parts separated by `` : '', on the left of the two points should be the name of the chemical reaction, and to its right the reaction in the form `` a A + b B + ... -$>$ c C + d D + ...'' where the low case letter represent numbers for the stoichiometric coefficients and the upper case letters are metabolites (molecules) names. The `` -$>$ '' can be substituted by `` $<$-$>$ '' in the case when the reaction is reversible. When defining the object one can also give an optional input, also a string, which will be used to indicate the pathway name of a particular reaction. For the metabolite names one can use spaces, but the \texttt{Enzyme} class will remove these spaces from the names. One can also use symbols, like ``+'', ``-'', etc, but being careful not to confuse them with the ``+'' sign indicating the interaction of different molecules.

Example,

\bv
>>> from PyNetMet.enzyme import *
>>> enz1 = Enzyme("reac1 : A + 2 B -> C + D")
\end{verbatim}
in this example one defines the variable \texttt{enz1} which contains the reaction whose name is reac1, where one molecule of A combines with two molecules of B to result in one molecule of C and one of D. One should always put spaces surrounding the ``:'', ``-$>$'' and ``+'' signs, otherwise the symbols might get confused with the metabolite names.

The representation of an enzyme object will be its initial string, but with numbers transformed to float type. Note the following examples:

\bv
>>> print enz1
reac1 : 1.0 A + 2.0 B  -> 1.0 C + 1.0 D
>>> enz2 = Enzyme("reac2 : A + 2 B -> C D")
>>> print enz2
reac2 : 1.0 A + 2.0 B  -> 1.0 CD
>>> enz3 = Enzyme("reac3 : A + 2 B -> C+ D")
>>> print enz3
reac3 : 1.0 A + 2.0 B  -> 1.0 C+D
>>> enz4 = Enzyme("reac4 : A + 2 B -> C+ + D")
>>> print enz4
reac3 : 1.0 A + 2.0 B  -> 1.0 C+ + 1.0 D
\end{verbatim}
from these examples one should note that it is possible to use spaces in the metabolite names, but the \texttt{Enzyme} class will remove these spaces from the names. One can also use symbols like ``+'' or ``-''.

Apart from the methods in Table \ref{tab2}, the class \texttt{Enzyme} has also a few underscore methods programmed: the \texttt{\_\_add\_\_}, \texttt{\_\_sub\_\_}, \texttt{\_\_rmul\_\_} and others, which define how \texttt{Enzyme} objects can be summed, subtracted, multiplied by constants, etc. The result of these mathematical operations with chemical reactions is rather intuitive and one can refer to the manual for a few examples.

\subsection{Network}

The \texttt{Network} class defines a graph and contains many algorithms for its analysis. It should be initiated with one obligatory input and an optional one. The input for this class is the $N\times N$ adjacency matrix for the network ($N$ is the number of nodes in the network) and the optional one a list with the node's names, if this is not given the nodes will be named with numbers from 0 to $N-1$. The adjacency matrix, M, is a list of $N$ elements, where each element is a list with $N$ elements, each element being 0 or 1. If $M[i][j]$ is 1, it means that node $i$ has a directed connection to node $j$. If the M matrix is symmetric, the network is undirected, meaning there is no distinction between a link from node $i$ to $j$ or from node $j$ to $i$. Otherwise, the network is interpreted as a directed graph, where the connections have an incoming and outgoing node.

From the mathematical point of view, a network is defined by a list of nodes and a list of edges. In order to represent such an object we started from the adjacency matrix $M$. This is the $N\times N$ matrix with zeros and ones, where $N$ is the number of nodes in the network and a directed edge exist coming from node $i$ to node $j$ if the element $M_{ij}$ of the matrix is 1. In this case of an undirected network, the total number of edges is half the total number of ones in the $M$ matrix. Another way to define a network is with a list of $N$ elements where each element of this list is a list of neighbors for a given node. In our class three such lists are created with the input $M$: \texttt{linksin}, \texttt{linksout} and \texttt{neigbs} for the list of incoming edges to a given node, outgoing ones and the total number of edges disregarding the directionality, respectively.

One can define a few attributes for each node. First, the node's degree is the number of connections it has to other nodes. In its calculation we do not consider the directionality of the connection, so what the class actually calculate in this attribute (\texttt{kis}) is the length of each element in the list \texttt{neigbs}.

Another attribute of a node is its clustering coefficient. It is defined by:

\be
C_i & = & \frac{2 E_i}{k_i(k_i-1)}
\ee
where $k_i$ is the degree of node $i$, and $E_i$ is the number of connections between the neighbors of node $i$. The average clustering of a network can be calculated straightforward:

\bv
>>> Cbar = sum(net.Cis)/net.nnodes
\end{verbatim}
given that the variable \texttt{net} is a \texttt{Network} object.

Next, we  define the topological overlap ($O_{ij}$) between two nodes according to \cite{nw3}:

\be
O_{ij} &=& \frac{C_{ij} + \left\{\bt{cl}1 & \textrm{, if $i$ connected to $j$} \\
                                         0 & \textrm{, otherwise}\et\right.}{\textrm{min}(n_i,n_j)}
\ee
where $C_{ij}$ is the number of common neighbors between nodes $i$ and $j$ and $\textrm{min}(n_i,n_j)$ is the minimum between the number of neighbors of nodes $i$ and $j$.

In \cite{nw3} a method for grouping the nodes in clusters is proposed basically by constructing a dendrogram (tree) with the values of the topological-overlap ($O_{ij}$). This tree can be constructed with the Kruskal algorithm, which is implemented in the Network class. Another interesting method for ordering the nodes is proposed in \cite{gwt}. Although this later method has many improvements with respect to the dendrogram one, it is based on a Monte-Carlo simulation and is computationally very costly. Here we propose yet a different method which is computationally more efficient and returns results at least as good as the dendrogram method.

The objective of the method is to reorder the nodes in the adjacency matrix (or the topological overlap one), such that nodes close to each other are correlated in the sense that they share neighbors which are also correlated among them, obtaining in this way an ordering where nodes belonging to common clusters are nearby each other. The algorithm follows the following steps:

\begin{enumerate}
\item[(1)] Choose any node $i$ to start with. Add it to the ordering.
\item[(2)] From node $i$, find the node $j$ for which $\chi^2_{ij}$ defined below is minimum:
\be
\chi^2_{ij} &=& \Sum_{k\in E^\prime} \frac{1}{\textrm{max}(0.00001,C_k)}\left(\frac{O_{ik}-O_{jk}}{O_{ik}+O_{jk}}\right)^2
\ee
where $E^\prime$ is the set of all nodes that have not been added to the ordering.
\item[(3)] Add node $j$ to the ordering.
\item[(4)] Set node $j$ as $i$ and repeat the process from step (2) until the set $E^\prime$ is empty. 
\end{enumerate}

The use of the function $\textrm{max}(0.00001,C_k)$ is to avoid a division by zero in the case that node $k$ has 0 clustering coefficient. This algorithm is implemented in the \texttt{Network} method \texttt{plot\_nCCs}. In the next section plots obtained from this method and with the dendrogram one for real metabolic models are shown.

\subsection{Metabolism}

This class defines an object with a full metabolic model. The metabolic model can be given as input in three different ways. By default one can use a single input which is a string containing the file name (with path) of a metabolic model in OptGene format, alternatively, one can use a file in SBML format and finally one can define lists containing reactions, constraints, external metabolites and objective function directly from the command line and use them as input for the class. So, this class works either as a parser for OptGene or SBML file formats or as a platform to construct new metabolic models from the beginning. 

This class has also the \texttt{dump} method, that allows one to write an output file with the stored model either in OptGene or SBML file formats. This resource allows the class to be used as a translator between OptGene and SBML file formats, for one can load the model in one format and dump it in the other format.

The main attribute from this class is its \texttt{enzymes} list, which contains all chemical reactions in the model. This list can be altered either directly (which is not advisable since other attributes of the class will not be automatically updated unless one calls the \texttt{calcs} method afterward), or by making use of the \texttt{bad\_reacs}, \texttt{add\_reacs} and \texttt{pop} methods. 

The use of this class together with the \texttt{Network} and \texttt{FBA} classes offers rich resources for an extensive analysis of any metabolic model.

\subsection{FBA}

The \texttt{FBA} class offers tools for performing flux simulations and analysis of a metabolic model. It has methods defined which are based on the FBA for studying essential reactions, sensibility of the objective function with respect to any given reaction, comparison of different realizations of the FBA, among others. 

To call this class one must give one obligatory input, which is a \texttt{Metabolism} object with a metabolic model. It can also receive two optional inputs with are the precision (\texttt{eps}, value under which a flux is considered zero, by default it is set to $10^{-10}$) and a choice of maximizing or minimizing the objective (the default choice is maximize).

This class has one underscore method, the \texttt{\_\_sub\_\_} which defines the subtraction of two FBAs. This method actually compares two different FBA outputs, returning a string with four columns, the first with the name of each reaction, the second and third with the flux of the reaction in each one of the FBAs and the fourth one with the relative difference in the fluxes ($100\%\frac{\nu_1-\nu_2}{\nu_1}$), in the case where the first flux is zero and the second is not, it returns the string ``NA'' in this column.


\section{Results and Discussion}

In this section we exemplify some uses of our tools by analyzing real metabolic models taken from the literature. 

We chose three models to work with, the first is the iSyn811 model of \syn \cite{bf3}. The second is the metabolic model iCM925 for the organism {\it Clostridium beijerinckii} NCIMB 8052 \cite{cbe} and last is the model iAK692 for {\it Spirulina platensis} C1 from \cite{spiru}. This last model comes in three different versions, we are using the first one. All these models are available from the journals as supplementary materials, the first one in OptGene format and the other two in SBML format. These have been downloaded and saved in a working folder and can be directly accessed by PyNetMet tools.

\bv
>>> from PyNetMet.metabolism import *
>>> syn=Metabolism("iSyn811.txt")
>>> cbe=Metabolism("iCM925.xml",filetype="sbml")
>>> ak=Metabolism("iAK692.xml",filetype="sbml")
\end{verbatim}

One might notice a discrepancy in the number of reactions and metabolites between the model reported in the literature and the one loaded by the tools. This is because the \texttt{Metabolism} class adds to the model transport reactions that are needed to perform the FBA. These added reactions are included in a pathway named \texttt{\_TRANSPORT\_}. The number of reactions in this pathway should equal the difference between the numbers reported in the literature and the actual number of reactions and metabolites in the loaded model.

\bv
>>> print cbe
# Reactions:957
# Metabolites:900
>>> print cbe.pathnames
['GPW', '_TRANSPORT_']
>>> print len(cbe.pathways[1])
19
\end{verbatim}
The iCM925 model, for instance, is reported to have 938 reactions and 881 metabolites, while the object cbe has 957 reactions and 900 metabolites. The difference (19) is the number of transport reactions in the \texttt{\_TRANSPORT\_} pathway.

First, let's plot a representation of the topological overlap of the nodes. For each model we make three plots, the first with the arbitrary order in which the nodes appear in the model, then with the nodes ordered by the Kruskal algorithm and finally with the algorithm described in subsection 2.2.

\bv
>>> syn.net.plot_matr(syn.net.nCCs, range(syn.nmets), output="plot_syn1.jpg")
>>> syn.net.kruskal(syn.net.nCCs, minimo=False)
>>> syn.net.plot_matr(syn.net.nCCs, syn.net.krusk_ord, output="plot_syn2.jpg")
>>> syn.net.plot_nCCs(output="plot_syn3.jpg")
>>> cbe.net.plot_matr(cbe.net.nCCs, range(cbe.nmets), output="plot_cbe1.jpg")
>>> cbe.net.kruskal(cbe.net.nCCs, minimo=False)
>>> cbe.net.plot_matr(cbe.net.nCCs, cbe.net.krusk_ord, output="plot_cbe2.jpg")
>>> cbe.net.plot_nCCs(output="plot_cbe3.jpg")
>>> ak.net.plot_matr(ak.net.nCCs, range(ak.nmets), output="plot_ak1.jpg")
>>> ak.net.kruskal(ak.net.nCCs, minimo=False)
>>> ak.net.plot_matr(ak.net.nCCs, ak.net.krusk_ord, output="plot_ak2.jpg")
>>> ak.net.plot_nCCs(output="plot_ak3.jpg")
\end{verbatim}
These commands should produce the nine plots shown in figure \ref{fig1}.

The average clustering for each network can easily be obtained:

\bv
>>> print sum(syn.net.Cis)/syn.net.nnodes
0.16599889162
>>> print sum(cbe.net.Cis)/cbe.net.nnodes
0.24542198734
>>> print sum(ak.net.Cis)/ak.net.nnodes
0.199707555775
\end{verbatim}

Other interesting analysis that can be made using the methods from the \texttt{Network} class are the search for disconnected components or the study of paths between the nodes of the metabolic network. Once the method \texttt{components} is called for a network, apart from the \texttt{disc\_comps} attribute, it automatically creates two new attributes, \texttt{dists} and \texttt{paths} that contain the shortest distances and paths in between any two nodes of the network.

\bv
>>> syn.net.components()
>>> print [len(ele) for ele in syn.net.disc_comps]
[976, 2, 5, 2, 2, 2]
\end{verbatim}
The attribute \texttt{disc\_comps} is a list with the list of nodes in each disconnected component of the network. In the above example we printed the number of nodes in each component, which shows us the giant component (976 metabolites) that comprises the metabolism, and 5 other components which are the result of reactions disconnected from the main metabolism and that could be removed from the metabolic model. The metabolism method \texttt{bad\_reacs} removes these reactions and also reactions where one product and one substrate only appear once in the whole metabolism, indicating that these reactions are also poorly connected to the main component.

For the other networks:
\bv
>>> cbe.net.components()
>>> print [len(ele) for ele in cbe.net.disc_comps]
[898, 2]
>>> ak.net.components()
>>> print [len(ele) for ele in ak.net.disc_comps]
[797, 5, 3, 3, 3, 2, 2, 3, 3, 3, 3, 3, 3, 3, 3]
\end{verbatim}

In the above examples the network under study is the one composed only by the metabolites in each metabolic model. One can chose to work with the bipartite network formed by metabolites and reactions. In the following examples we build this network in order to study paths between metabolites.

\bv
>>> import sys # For setting a new recursion limit for recursive functions
>>> sys.setrecursionlimit(10000)
>>> [Mreac,names] = syn.M_matrix_reacs()
>>> net = Network(Mreac, names)
>>> iglu = names.index("alpha-D-glucose")
>>> ipyr = names.index("pyruvate")
>>> [paths, dists] = net.calc_dist_wp(net.linksout, iglu)
>>> print [names[ii] for ii in paths[ipyr]]
['alpha-D-glucose', '2.7.1.2b', 'ADP', '2.7.1.40a', 'pyruvate']
>>> print paths[ipyr]
[4, 990, 3, 1003, 21]
\end{verbatim}
In this example we calculated the shortest path from glucose to pyruvate: it goes through reaction 2.7.1.2b, which has ADP as product and ADP is substrate in reaction 2.7.1.40a that produces pyruvate. Note that if one prints \texttt{paths[ipyr]} the numbers that one sees are [4, 990, 3, 1003, 21]. The numbers 3, 4 and 21 correspond to the positions of ADP, alpha-D-glucose and pyruvate in the list \texttt{syf.metabol}, but the numbers that correspond to reactions 2.7.1.2b and 2.7.1.40a in the list \texttt{syf.enzymes} are not 990 and 1003, but instead 1 (990-\texttt{syf.nmets}) and 14 (1003-\texttt{syf.nmets}).

Metabolites that could not be reached from glucose are marked with the symbol ``X'' in the dists list:

\bv
>>> dists.count("X")
221
>>> ndist = filter(lambda x:x != "X", dists)
>>> print 1.*sum(ndist)/len(ndist)
5.78828081813
>>> print max(ndist)
20
>>> dists.count(20)
1
>>> names[dists.index(20)]
'Astxbm'
\end{verbatim}
Here we see that 221 metabolites could not be reached from glucose. From those that could be reached, the average shortest path is around 5.788 and the furthest metabolite reached by glucose is Astxbm which is 20 nodes away.

Apart from the network analysis of the models, one can use the methods in class FBA. Just by calling the class with a metabolic model as input one can directly obtain the FBA result by printing the class object. The result is a string listing the reaction names and their respective fluxes ordered, by default, according to these fluxes.

\bv
>>> from PyNetMet.fba import *
>>> fba_cbe=FBA(cbe)
>>> print fba_cbe
                                     ...
                          < ... lots of output ... >
                                     ...
                        R_GAPD ---->  5.012953336  
                       R_FDXNH ---->  6.963928797  
                        R_H2ex ---->  6.963928797  
                     R_ex_h2_e ---->  6.963928797  

Flux on objective                   : 0.119427648 
Reactions with flux    (flux>eps)   : 284 
Reactions without flux (flux<eps)   : 673 
Solution status: Optimal
\end{verbatim}

The \texttt{FBA} objects have the method \texttt{\_\_sub\_\_} defined, which allows a comparison between two realizations of a FBA. As an example on how it works, let's compare the metabolism of \syn when optimizing its growth and when optimizing hydrogen production for a fixed value of growth.

\bv
>>> fba_syn1=FBA(syn)
>>> print fba_syn1.Z
0.0895186102158
>>> gro = fba_syn1.Z
>>> syn.constr[syn.dic_enzs["_Growth"]] = (0.95*gro, 0.95*gro)
>>> syn.obj = [("_H2","1")]
>>> fba_syn2 = FBA(syn)
>>> print >>open("diff.txt","w"), fba_syn1-fba_syn2
\end{verbatim}

In this series of commands we create the first FBA where the growth (reaction named ``\_Growth'') of \syn is optimized. We then use the metabolic model to create a second FBA where the growth is fixed to 95\% of its optimized value and then optimize the production of hydrogen (reaction named ``\_H2''). The last command will create a file called diff.txt (additional file 2) where one can see the comparison between this two states of the metabolism. This file shows four columns, the first one is the name of each reaction, in the second and third one can find the values for the fluxes in each FBA, respectively. The fourth column shows the absolute value of the relative change in percentage ($100\%\left|\frac{\nu_1-\nu_2}{\nu_1}\right|$). If the original flux was zero ($\nu_1=0$) it will return ``NA'' in this column. By default it sorts the reaction by its difference value, so the first reactions listed on the file will have no difference in their flux and the reactions in the end of it will be the affected ones. One can clearly see that the most affected reactions are those related with the \syn hydrogenase.

Another straight forward method to analyze a FBA is the \texttt{essential} method, which checks if a reaction is essential for producing flux in the objective function. When called, it returns one Boolean value stating if the reaction is essential or not and a second value which informs of the relative change in the objective flux with the input reaction removed.

\bv
>>> print syn.enzymes[926]
_H2CO3transport : 1.0 H2CO3_extrac  <-> 1.0 H2CO3
>>> print fba_syn1.essential(926)
[False, 0.50000000000000111]
\end{verbatim}
This tells us that the transport of carbonic acid is not essential for the growth of the cell, but its removal reduces the growth by 50\%. We can also count the total number of essential reactions in each model:

\bv
>>> print sum([fba_cbe.essential(ii)[0] for ii in xrange(cbe.nreacs)])
166
>>> print sum([fba_syn1.essential(ii)[0] for ii in xrange(syn.nreacs)])
221
>>> fba_ak=FBA(ak)
Warning: repeated index in Stoichiometric matrix. Look into reaction 746. FBA might not be correct.
Warning: repeated index in Stoichiometric matrix. Look into reaction 748. FBA might not be correct.
>>> print ak.enzymes[746]
R_MotexX : 1.0 M_Mo_e  <-> 1.0 M_Mo_e 
>>> print ak.enzymes[748]
R_NatexX : 1.0 M_Na_e  <-> 1.0 M_Na_e 
>>> print sum([fba_ak.essential(ii)[0] for ii in xrange(ak.nreacs)])
249
\end{verbatim}
So, the iCM925 model has 166 essential reactions, the iSyn811 221 and the iAK692 has 249. One also sees that the FBA class recognizes two problematic reactions in iAK692 model, in this case, where a metabolite is connected to itself by the reaction.

The \texttt{shadow} and \texttt{max\_min} methods work in a similar way. In each method one has to use as input an integer indicating a reaction number. The shadow method has two other optional inputs, the first one indicates the change in the original flux in order to calculate the derivative (the result should be independent of this choice, since the problem is linear) and the second indicates if one wishes the relative change or the absolute change. 

The method \texttt{max\_min} has the \texttt{fixobj} optional input. Its algorithm fixes the flux in the objective function to its original value times the value in \texttt{fixobj} (this should always be a value between 0 and 1). Then it sets the input reaction as objective and minimizes it, then maximizes it in its natural direction ($S\rightarrow P$) and then again minimizes it and maximizes it in the reversed direction ($S\leftarrow P$). It returns a two element list, each element is a tuple with the value of the flux in the reaction minimized and maximized in the direct direction and in the reversed direction, respectively. If the reaction is irreversible or there was no feasible solution for some optimization, it returns the string ``X''.

\bv
>>> print fba_syn1.shadow(926)
0.5
>>> print fba_syn1.shadow(926, relat=False)
0.0263290030046
>>> print fba_syn1.max_min(926,fixobj=0.5)
[(0.0, 671.92509141136259), (0.0, 0.0)]
>>> print fba_syn1.max_min(926,fixobj=0.6)
[(0.34000000000011554, 670.31010969363479), ('X', 'X')]
\end{verbatim}
As we saw before, the transport of carbonic acid can be removed at cost of reducing by a factor two the growth in the iSyn811 model. So, calculating its maximal and minimal flux if we fix the growth to half its maximum value, we find that we don't need the reaction (one is able to minimize it to zero). But, if we fix the growth to 60\% of its maximal value, the flux in the transport of carbonic acid must be at least equal to 0.34 and in this case the reaction cannot occur in the reversed direction (which is indicated by the X's in the second list).

All these examples do not intend to exhaust the uses of the PyNetMet tools and their functionality, but should be enough to illustrate their potentiality.


\section{Conclusions}

We have presented the PyNetMet package which contains four classes (\texttt{Enzyme}, \texttt{Network}, \texttt{Metabolism} and \texttt{FBA}) intended to facilitate the analysis, work, curation and construction of networks and metabolic models. These tools allow one to work with metabolic models in either standard (OptGene and SBML) and to easily convert one to another. The \texttt{Metabolism} class can be used as a platform to produce variants of any model ({\it in silico} mutants) by producing knock-ins or knock-outs with the \texttt{add\_reacs} and \texttt{pop} methods, respectively and studying its effects straightforwardly with the class \texttt{FBA}.

These tools are in the format of Python modules, which allow the researcher to integrate them with any other Python resource available. They are also open-source and free software which allows one to develop new tools using these as building blocks.

This work also provides complementary examples (to the ones found in the manual) for uses of these tools with real published metabolic models.

The authors present these tools in the hope that the scientific community will find them useful in their researches and might even extend and use them in yet new useful tools contributing even further the development in this field.

\bigskip

\section*{Author's contributions}

D.~Gamermann programmed the classes and together with R. Jaime debugged the FBA class. A.~Montagud and J.~Triana analyzed data and the results and these authors contributed to the writing of the manuscript. P~.F~. de C\'ordoba and J.~Urchuegu\'ia conceived and funded the study. All authors read and approved the manuscript.

\section*{Acknowledgements}
  \ifthenelse{\boolean{publ}}{\small}{}
  This work has been funded by the MICINN TIN2009-12359 project ArtBioCom from the Spanish Ministerio de Educaci\'on y Ciencia and FP7-ENERGY-2012-1-2STAGE (Project number 308518) CyanoFactory from the EU.


\newpage
{\ifthenelse{\boolean{publ}}{\footnotesize}{\small}
 \bibliographystyle{bmc_article}  
  \bibliography{bmc_article} }     


\ifthenelse{\boolean{publ}}{\end{multicols}}{}


\section*{Figures}

\subsection{Figure 1 - plot\_syn1, plot\_syn2, plot\_syn3, plot\_cbe1, plot\_cbe2, plot\_cbe3, plot\_ak1, plot\_ak2 and plot\_ak3}\label{fig1}

\mbox{
\bt{ccc}
\ig{0.3}{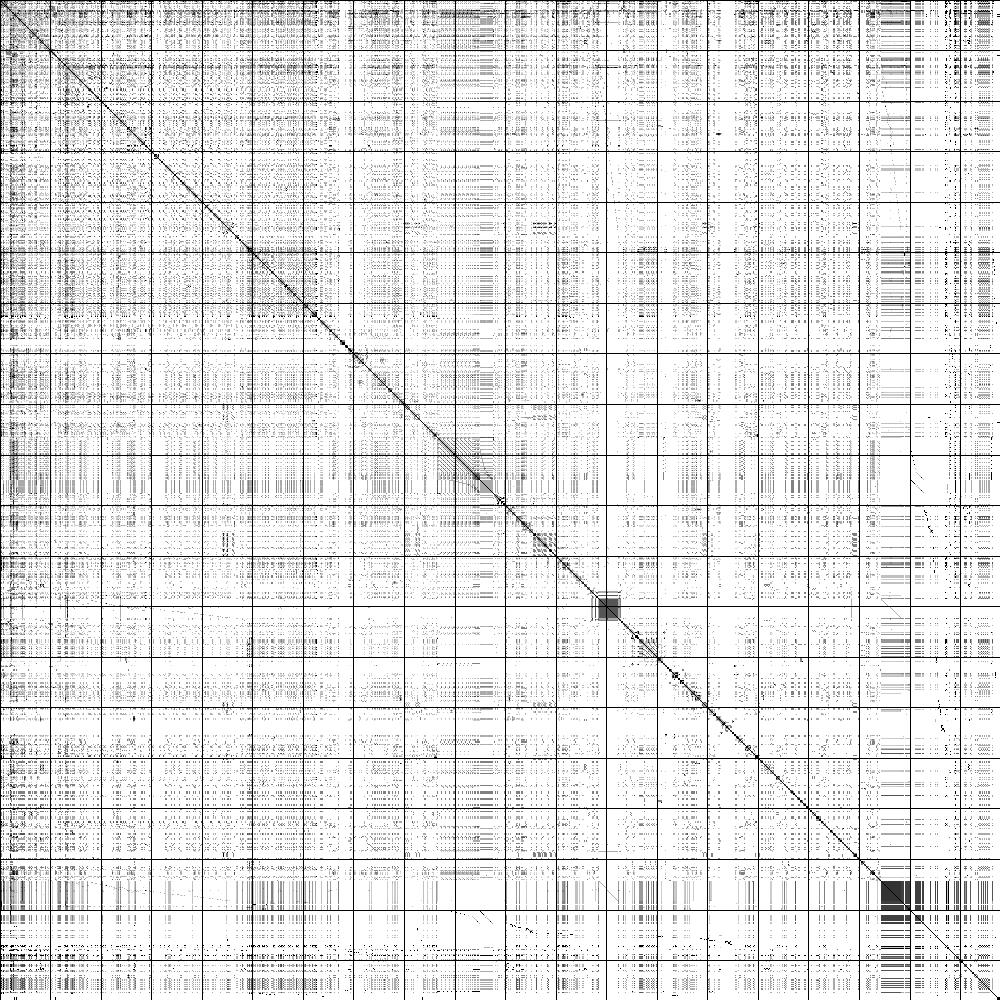} & \ig{0.3}{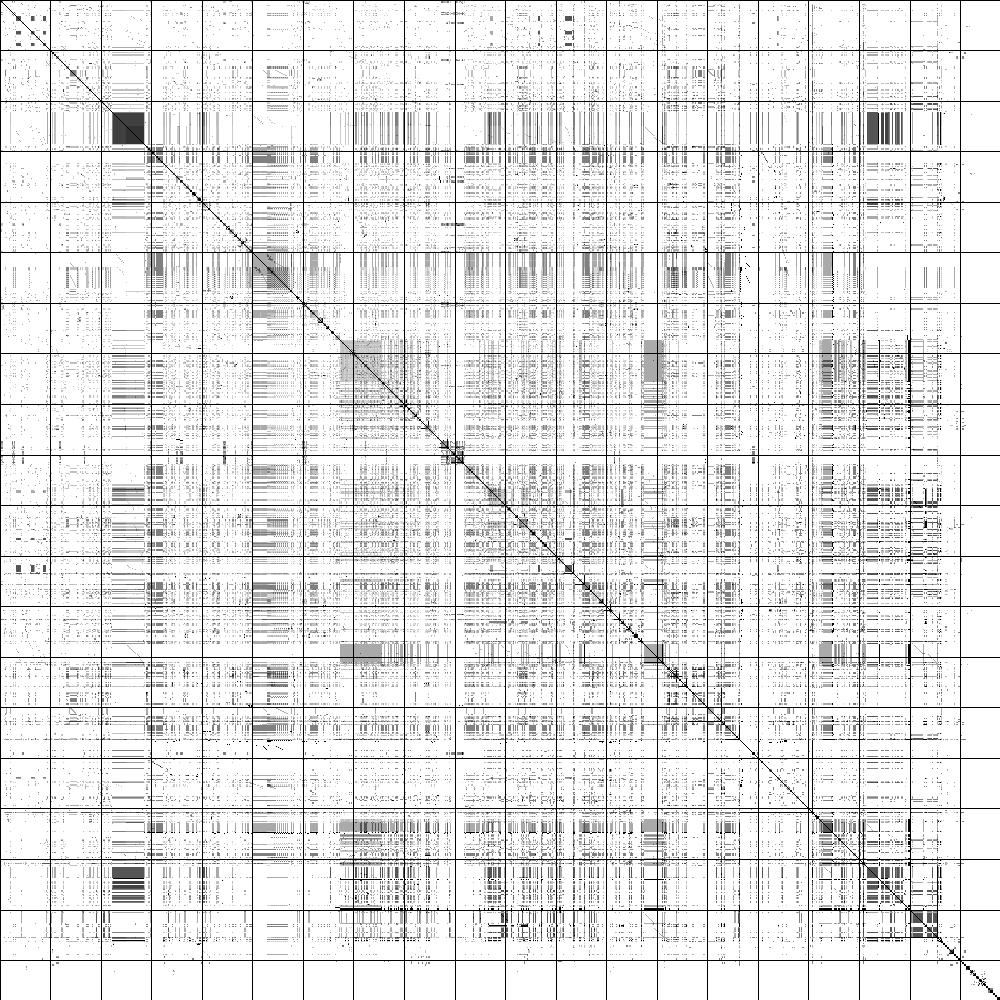} & \ig{0.3}{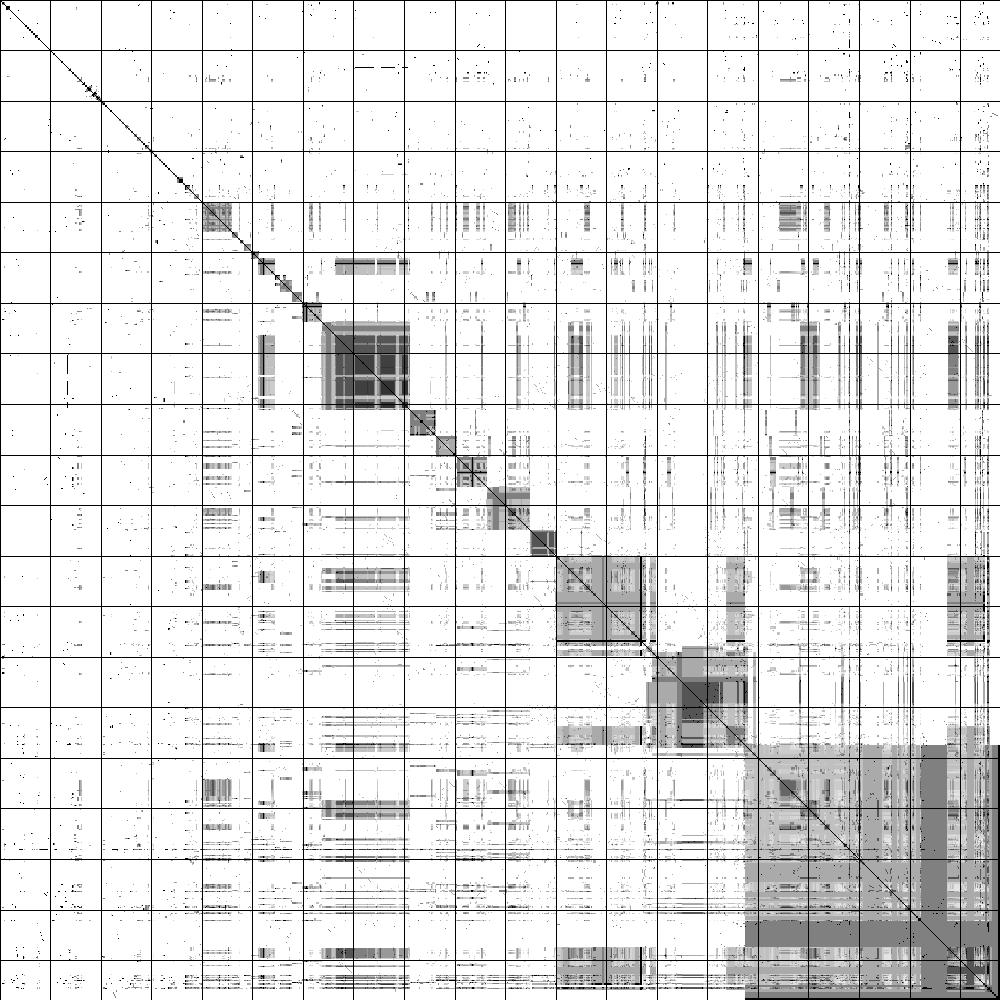} \\
\ig{0.3}{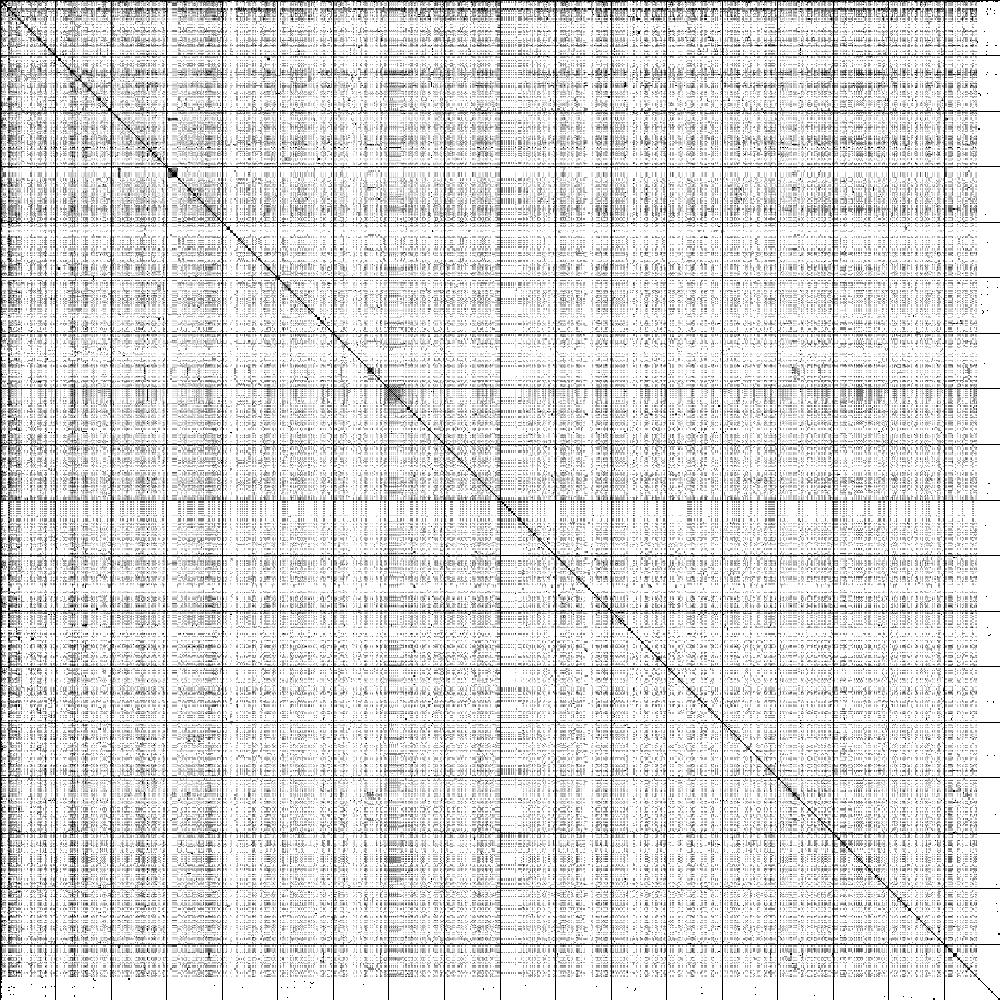} & \ig{0.3}{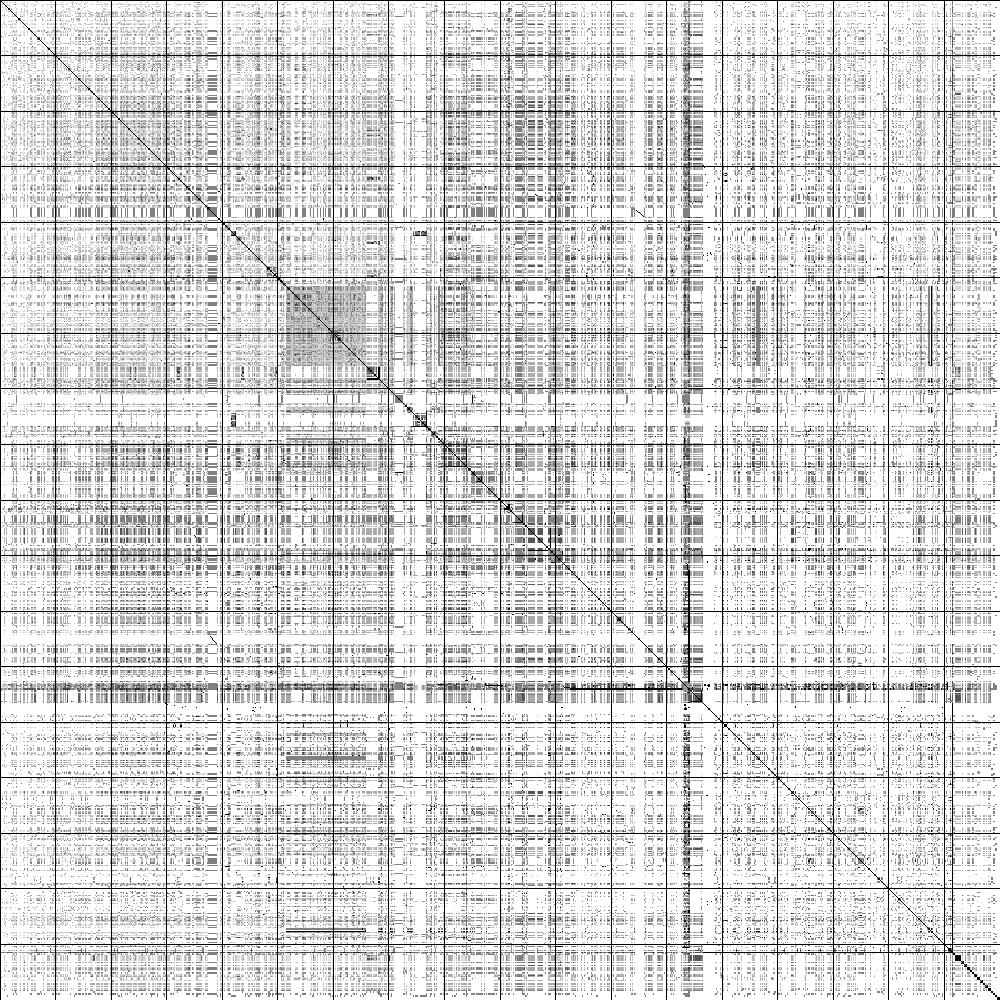} & \ig{0.3}{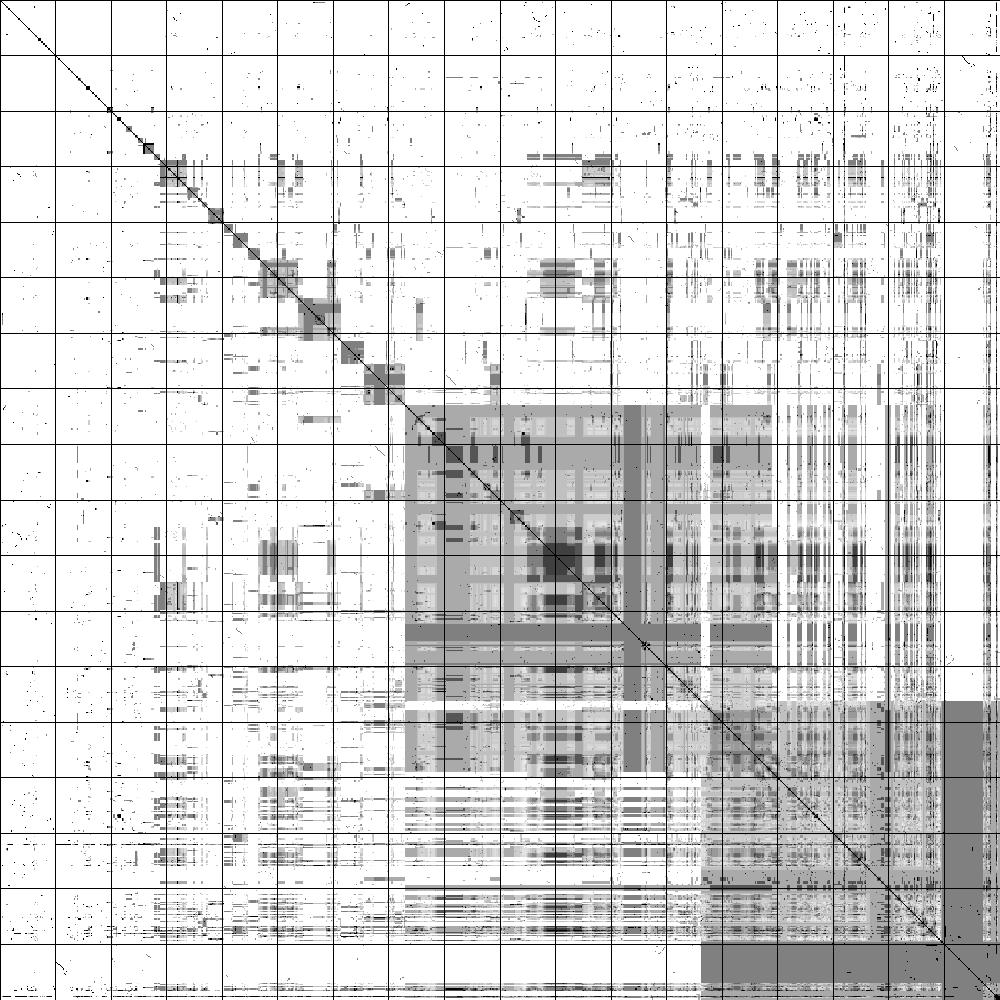} \\
\ig{0.3}{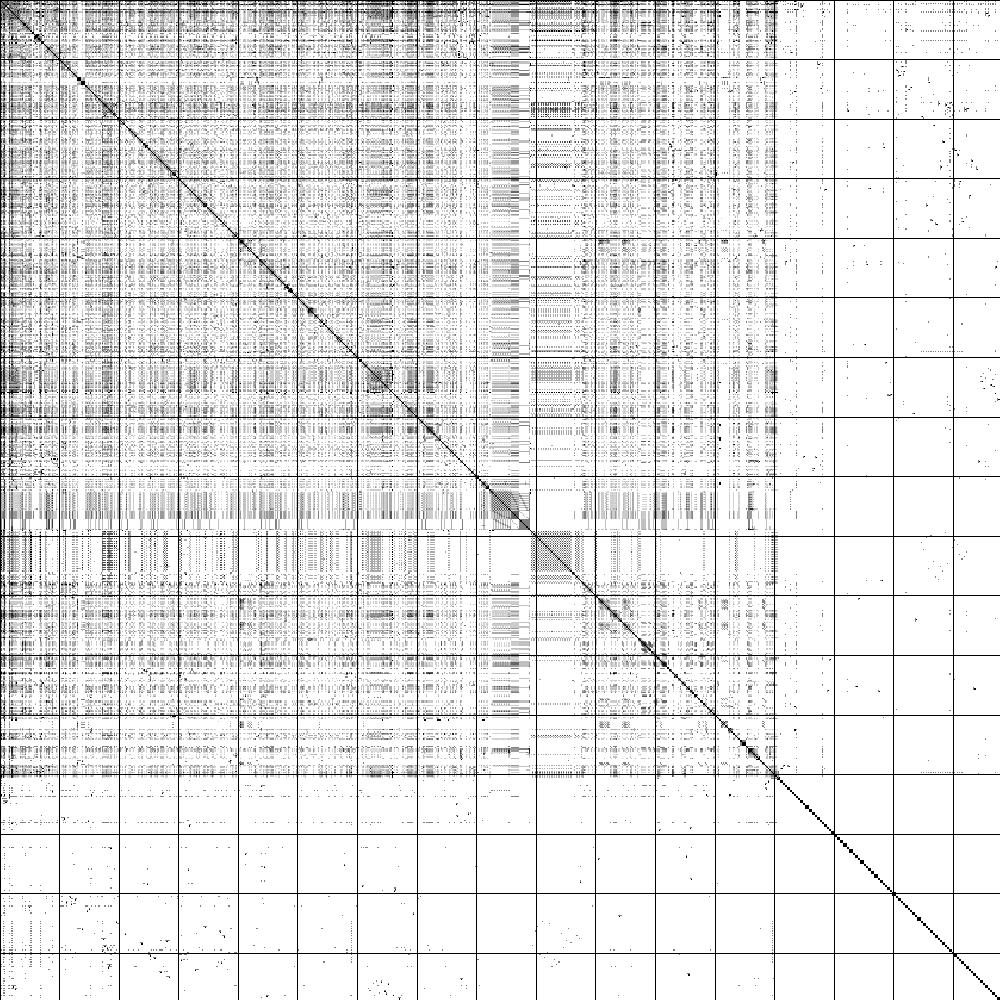} & \ig{0.3}{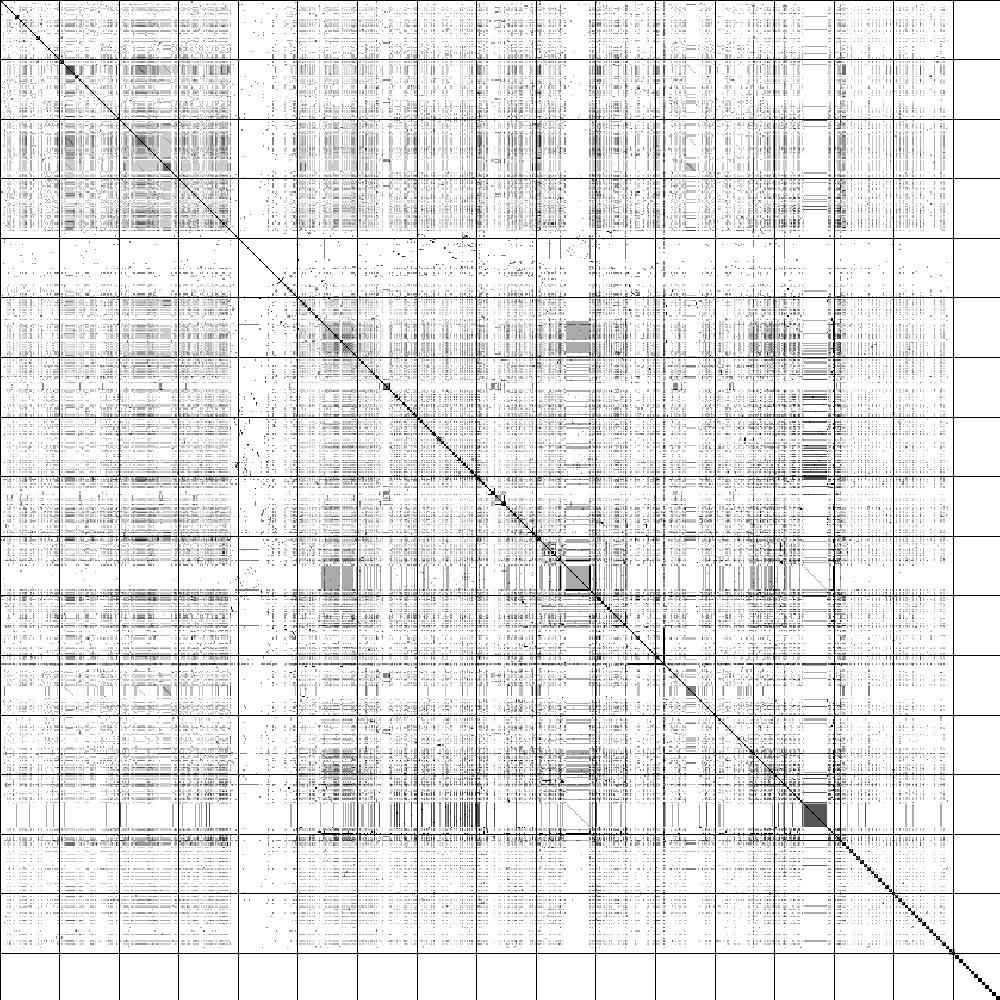} & \ig{0.3}{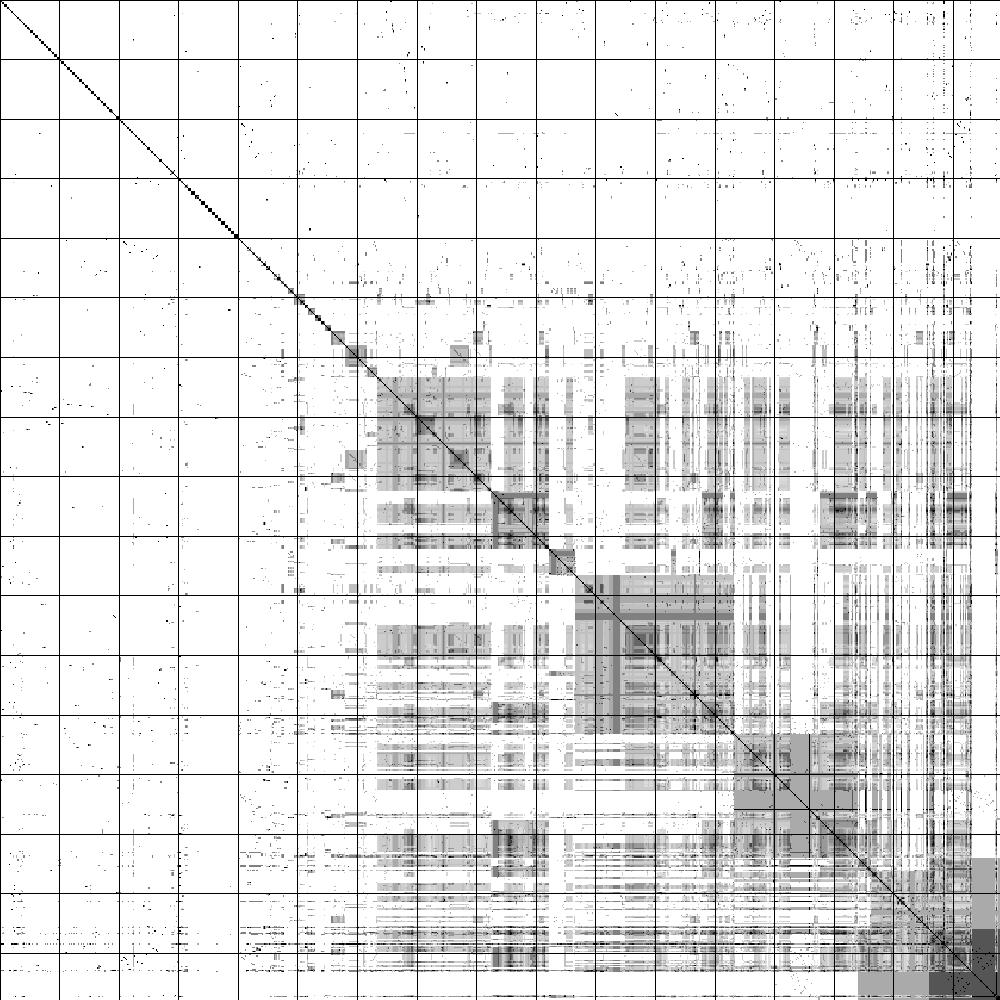} 
\et
}

Plots for the topological overlap of metabolites. The first, second and third rows refer to the plots obtained by the three analyzed models, iSyn811, iCM925 and iak692, respectively. The plots in the first column are for an arbitrary ordering of the metabolites, in the second column an ordering is obtained via the Kruskal algorithm and in the third column the ordering is obtained by the algorithm implemented in the \texttt{plot\_nCCs} method of the \texttt{Network} class.


\section*{Tables}

\subsection{Table 1 - Attributes of the classes}\label{tab1}

\mbox{\scriptsize
\bt{lll}
Class & Attribute & Brief description \\
\hline
Enzyme &  name & Name of the reaction. \\
Enzyme &  pathway & The pathway to which the reaction belongs. \\
Enzyme &  reversible & Indicates if the reaction is reversible. \\
Enzyme &  Nsubstrates & Number of substrates. \\
Enzyme &  Nproducts & Number of products. \\
Enzyme &  metabolites & List of all metabolites in the reactions. \\
Enzyme &  substrates & List of substrates. \\
Enzyme &  products & List of products. \\
Enzyme &  issues & Indicates possible issues in the reaction. \\
Enzyme &  issues\_info & List of issues. \\
Enzyme &  stoic & List of stoichiometric coefficients. \\
Enzyme &  tup & Tuple with the number of substrates and products. \\
Network & nnodes & Number of nodes in the network. \\
Network & nodesnames & List with the names of the nodes. \\
Network & directed & Indicates if the network is directed or not. \\
Network & links & List of tuples, indicating all edges in the graph. \\
Network & nlinks & Total number of directed connections in the network. \\
Network & linksin & List of edges coming from a given node. \\
Network & linksout & List of edges going to a given node. \\
Network & neigbs & List of all nodes with connection to or from a given node. \\
Network & kis & List with the degree of each node. \\
Network & Cis & List with the clustering coefficients of each node. \\
Network & CCs & Matrix with lists of common neighbors of two nodes. \\
Network & uCCs & Matrix with lists of all neighbors of two nodes. \\
Network & nCCs & Matrix with the topological overlap of two nodes. \\
Network & weight & List with the average of each row of matrix nCCs. \\
Network & sd\_wei & List with the standard deviation for each element of weight.\\
Metabolism & file\_name & Name of input file (or model). \\
Metabolism & enzymes & List of all reactions in the model (Enzyme objects). \\
Metabolism & dic\_enzs & Association between reaction name and position in the lists. \\
Metabolism & nreacs & Total number of reactions. \\
Metabolism & reac\_irr & List of positions of the irreversible reactions. \\
Metabolism & reac\_rev & List of positions of the reversible reactions. \\
Metabolism & nreac\_irr & Number of irreversible reactions. \\
Metabolism & nreac\_rev & Number of reversible reactions. \\
Metabolism & mets & List of all metabolites in the model. \\
Metabolism & dic\_mets & Association of metabolite name and position in the lists. \\
Metabolism & nmets & Total number of metabolites. \\
Metabolism & pathnames & List of comments found in the model file. \\ 
Metabolism & pathways & List with lists of reactions per pathway. \\
Metabolism & reac\_per\_met & List of lists of reactions where each metabolite appears. \\
Metabolism & reacs\_per\_met & List with the number of reactions where each metabolite appears. \\
Metabolism & M & The adjacency matrix for the metabolites. \\
Metabolism & net & The metabolite's network (Network class object for the above adjacency matrix). \\
Metabolism & reactions & Raw data in OptGene format for the reactions.\\
Metabolism & transport & List of all transport reactions. \\
Metabolism & external & Raw data in OptGene format for the external metabolites. \\
Metabolism & external\_in & List of metabolites that can come inside the cell from the outside. \\
Metabolism & external\_out & List of metabolites that the cell transports to the outside. \\
Metabolism & constrains & Raw data in OptGene format for the constraints. \\
Metabolism & constr & List with constraints for each reaction. \\
Metabolism & objective & Raw data in OptGene format for the objective (for FBA optimization). \\
Metabolism & obj & List for the objective function. \\
FBA & reacs & List with the reactions (Enzyme objects). \\
FBA & nreacs & Total number of reactions. \\
FBA & reac\_names & List with the names of all reactions. \\
FBA & mets & List with all metabolites. \\
FBA & nmets & Total number of metabolites. \\
FBA & ext\_in & List of external metabolites that can enter the cell. \\
FBA & ext\_out & List of metabolites that can leave the cell. \\
FBA & Mstoic & List of tuples with the non-zero elements of the stoichiometric matrix. \\ 
FBA & constr & List of tuples with the constrains to be applied. \\ 
FBA & flux & List with the fluxes for all reaction from the last optimization. \\
FBA & Z & Value of the flux in the objective. \\
FBA & obj & List of tuples for the objective. \\ 
FBA & eps & Precision (value under which a flux is considered to be zero). \\
FBA & lp & Linear problem (Pyglpk object).
\et
}

\subsection{Table 2 - Methods of the classes}\label{tab2}

\mbox{\footnotesize
\bt{lll}
Class & Method & Brief description \\
\hline
Enzyme & connects & Checks if two metabolites are connected by the reaction. \\
Enzyme & copy & Returns a copy of the reaction. \\
Enzyme & has\_metabol & Checks if a metabolite appears in the reaction. \\
Enzyme & has\_product & Checks if a metabolite is product of the reaction. \\
Enzyme & has\_product\_rev & Checks if a metabolite is product (or substrate, if reversible) of the reaction. \\
Enzyme & has\_substrate & Checks if a metabolite is substrate of the reaction. \\
Enzyme & has\_substrate\_rev & Checks if a metabolite is substrate (or product, if reversible) of the reaction. \\
Enzyme & make\_irr & Returns the irreversible version of the reaction. \\
Enzyme & make\_rev & Returns the reversible version of the reaction. \\
Enzyme & pop & Removes a metabolite from the reaction. \\
Enzyme & rev\_reac & Returns the reaction in the reversed order (changes substrates and products). \\
Enzyme & stoic\_n & Returns the stoichiometric coefficient of a metabolite. \\
Network & plot\_nCCs & Orders and plots the topological overlap for the nodes. \\
Network & plot\_matr & Plots a given matrix with a given ordering. \\
Network & kruskal & Solves the Kruskal algorithm for a given matrix. \\
Network & calc\_all\_dists & Calculates distances of all nodes to all others.\\
Network & calc\_all\_dists\_wp & Same as calc\_all\_dists, but returning also the paths. \\
Network & calc\_dists & Same as calc\_all\_dists but for a single node.\\
Network & calc\_dists\_wp & Same as calc\_dists, but returning also the paths.\\
Network & components & Searches all disconnected components of the network. \\
Metabolism & calcs & Calculates all attributes based in the enzymes list. \\
Metabolism & add\_reacs & Adds reactions to the metabolism. \\
Metabolism & pop & Removes a single reaction from the metabolism. \\
Metabolism & dump & Writes an output file with the model. \\
Metabolism & M\_matrix & Returns the adjacency matrix for the metabolites network.\\
Metabolism & M\_matrix\_reacs & Returns the bipartite adjacency matrix of metabolites and reactions.\\
Metabolism & bad\_reacs & Removes reactions belonging to disconnected components of the network. \\
Metabolism & write\_log & Write a output file with information on the model.\\
FBA & print\_flux & Returns the flux of a single reaction.\\
FBA & fba & Prepares and performs the FBA. \\
FBA & shadow & Calculates the derivative (sensibility) of a given reaction.\\
FBA & essential & Tests whether a reaction is essential for producing flux in the objective.\\
FBA & max\_min & Returns the maximum and minimum flux of a reaction for a fixed objective value.
\et
}


\section*{Additional Files}

  \subsection*{Additional file 1 - PyNetMet\_manual.pdf}

Manual for the PyNetMet package.

  \subsection*{Additional file 2 - diff.txt}

Differences in the fluxes of the reactions of the iSyn811 model when optimizing growth or h2.

\end{bmcformat}

\end{document}